\documentclass[10pt,twocolumn,oneside,a4paper,final]{IEEEtran}
\usepackage{diagbox}
\usepackage{graphicx}
\usepackage{psfrag,amsbsy,pstricks,amsmath,amsfonts,amssymb}
\usepackage{algpseudocode}
\usepackage[numbers,sort&compress]{natbib}
\usepackage{epstopdf}
\usepackage{bm}
\usepackage[normalem]{ulem}
\usepackage{xcolor}

\usepackage{booktabs}
\usepackage{glossaries}
\newacronym{AD}{AD}{angle-delay}
\newacronym{AWGN}{AWGN}{additive white Gaussian
noise}

\newacronym{BS}{BS}{base station}
\newacronym{B-TSGM}{B-TSGM}{Bernoulli two state Gaussian mixture}

\newacronym{CE}{CE}{channel estimation}
\newacronym{DFT}{DFT}{discrete Fourier transform}

\newacronym{EM}{EM}{expectation maximization}

\newacronym{HMP}{HMP}{hybrid message passing}

\newacronym{PDF}{PDF}{probability density function}
\newacronym{PDFT-RP}{PDFT-RP}{partial DFT random permutation}

\newacronym{MC}{MC}{Markov chain}
\newacronym{MIMO-OFDM}{MIMO-OFDM}{multiple-input, multiple-output orthogonal frequency-division multiplexing}

\newacronym{OFDM}{OFDM}{orthogonal frequency-division multiplexing}

\newacronym{SCM}{SCM}{spatial channel model}

\newacronym{ULA}{ULA}{uniform linear array}
\newacronym{PDMP}{PDMP}{peak detection-message passing}

\usepackage[linesnumbered,ruled,vlined]{algorithm2e}
\SetKwInput{KwInput}{Input}                
\SetKwInput{KwOutput}{Output}              
\usepackage[caption=false]{subfig}
\usepackage{float}

\makeatletter 

\usepackage{tikz}
    \usetikzlibrary{shapes.arrows}
    \usetikzlibrary{matrix}

\usepackage{pgfplots}
\pgfplotsset{
compat=1.3,
legend style={font=\footnotesize, fill opacity=0.7,  draw opacity=1, text opacity=1, draw=white!15!black, legend cell align=left, align=left}, 
width=6cm, 
height=3cm,
yminorticks=false,
xminorticks=false,
title style={font=\scriptsize},
tick style={color=black},
tick label style={font=\scriptsize},
label style={font=\footnotesize},
grid style={line width=.1pt, draw=gray!20},
major grid style={line width=.1pt,draw=gray!20},
}
\usepgfplotslibrary{colorbrewer}
\usepgfplotslibrary{groupplots}

\definecolor{gold}{rgb}{201,166,0}
\definecolor{indigo}{rgb}{0,0,255}
\def \fwidth{.255\columnwidth}
\def \fwidthq{.20\columnwidth}

\newcommand{\T}{^{\mathsf{T}}}

\newcommand{\CN}{\mathcal{CN}}

\begin{document}
\title{Multi-dimensional Parameter Estimation in  RIS-aided MU-MIMO-OFDM Channels}
\author{Linlin~Mo, Yi~Song,   Fabio~Saggese,
Xinhua~Lu,
Zhongyong~Wang, and Petar~Popovski,~\IEEEmembership{Fellow,~IEEE}   \vspace{-0.7cm}

\thanks{L. Mo and X. Lu are with the
Academy for Electronic Information Discipline Studies, Nanyang Institute of Technology, 473000  Nanyang, China (linlinmoie@outlook.com; ieluxinhua@sina.com). 
Y. Song is with the Key Laboratory of Grain Information Processing and Control, Ministry of Education, Henan Engineering Research Center of Grain Condition Intelligent Detection and Application, Henan University of Technology, 450001 Zhengzhou, China (songyi@haut.edu.cn); Z. Wang is with the School of  Electrical and  Information Engineering, Zhengzhou University, 450001 Zhengzhou, China (zywangzzu@gmail.com). 
F. Saggese is with Dep. of Information Eng., Univ. of Pisa (fabio.saggese@ing.unipi.it) and P. Popovski is with the Department of Electronic Systems, Aalborg University (petarp@es.aau.dk).
 (Corresponding author: Xinhua Lu.)
}}

\maketitle
\begin{abstract}
We address the channel estimation (CE) problem in reconfigurable intelligent surface (RIS) aided orthogonal frequency-division multiplexing (OFDM) systems by proposing a dual-structure and multi-dimensional transformations (DS-MDT)  algorithm. 
The proposed approach leverages the  dual-structure features of the channel parameters to assist users experiencing weaker channel conditions, thereby enhancing CE performance.
Moreover, given that the channel parameters are distributed across multiple dimensions of the received tensor, the proposed algorithm employs multi-dimensional transformations to  isolate and extract distinct parameters. 
The numerical results demonstrate the proposed algorithm reduces the normalized mean square error (NMSE) by up to 10 dB while maintaining lower complexity compared to state-of-the-art methods.
\end{abstract}
\begin{IEEEkeywords}
Channel estimation, reconfigurable intelligent surface, broadband millimeter-wave, tensor. 
\end{IEEEkeywords}

\IEEEpeerreviewmaketitle

\section{Introduction}\label{Sec:intro}
Reconfigurable intelligent surfaces (RIS), as a transformative technology for the next generation of wireless communication,  have been extensively investigated  due to their  capability to dynamically manipulate the wireless propagation environment,  thereby substantially improving  communication performance~\cite{RISfuture}. 
Accurate channel state information (CSI) is indispensable for realizing the full potential of RIS. 
Consequently, channel estimation (CE) has been extensively studied.
CE under Doppler effects is addressed in~\cite{10925440}, and mutual coupling in~\cite{11176921}.
Additionally, addressing fabrication tolerances/calibration also remains a critical issue in RIS-aided CE~\cite{10341304}.
{
Beyond these impairments, a fundamental challenge persists in orthogonal frequency-division multiplexing (OFDM) systems~\cite{SCPD}: the coupling of massive reflecting elements and subcarriers induces high-dimensional system models, significantly complicating CE.
To concentrate on this structural challenge, we assume a quasi-static environment, abstracting away  practical impairments~\cite{9844229,SCPD, PDMP,8879620}.
}

Recent research efforts have explored tensor techniques to improve the CE performance~\cite{9844229,SCPD, PDMP}. 
The authors in~\cite{9844229} developed a novel CE method employing sparsity-structured tensor factorization, integrating compressive sensing principles with tensor decomposition mechanisms to achieve accurate channel recovery while maintaining minimized training requirements.
The work in~\cite{SCPD} proposed a structured tensor decomposition framework that exploits inherent sparse scattering characteristics through canonical polyadic decomposition (CPD) to enable efficient channel parameter estimation.
In~\cite{PDMP}, the authors proposed a direct tensor-based CE algorithm, where the multi-dimensional structure of the tensor is used.
However, two critical issues require further consideration.
Unlike traditional  systems where channel parameters of different users are mutually independent, the cascaded channel parameters of RIS-aided  systems exhibit dual-structure features.
Moreover, the existing tensor methods are not directly applicable with high-resolution parameter estimation techniques, e.g., multiple signal classification (MUSIC) algorithm.

To solve these challenges, we develop a dual-structure and multi-dimensional transformation (DS-MDT) algorithm.
The main contributions  can be summarized as follows: 
$i$) We reveal the dual-structure  features of the  cascaded channel parameters, namely, \emph{common} and \emph{offset} features, generated by the common RIS-BS channel experienced by all the UEs. 
$ii$) We show that multi-dimensional channel parameters (including angle, delay, and gain) are contained in distinct dimensions of the receive tensor. 
$iii$) We employ the multi-dimensional transformation (MDT) method to separate the respective dimensions, and perform the MUSIC algorithm to estimate the channel parameters.
The numerical results show the superior performance and lower complexity of the proposed algorithm.

\textit{Notation:} 
Scalars, vectors, matrices, and tensors are denoted by lowercase $a$, boldface lowercase  $\bm a$, boldface
capitals $\bm A$ and calligraphic $\cal A$ letters, respectively.
$\circ$, $\otimes$, $\odot$, $\left[\kern-0.15em\left[ \cdot \right]\kern-0.15em\right]$ and $\|\cdot\|$ 
denote the outer, Kronecker, Khatri–Rao and Kruskal  products, and the Euclidean norm. 
$(\cdot)\T$ and $(\cdot)^\dagger$  
represent  transposition and  pseudo inverse. 
$a_{i}$,  $a_{i,j}$ and $\bm A_{(:,a:b)}$ denote the $i$-th element of $\bm{a}$, the $(i,j)$-th element of $\bm A$ and columns $a$–$b$ of  $\bm A$.    $\CN(\mu_x,v_x)$ denotes the complex Gaussian distribution with mean $\mu_x$ and variance $v_x$; uniform distribution on $[a,b]$ is denoted by ${\cal{U}} [a,b]$. The estimation of $x$ is $\hat{x}$.

\section{System Model} 
We focus on the uplink CE in an RIS-aided millimeter-wave (mmWave) OFDM system. 
The BS employs a uniform linear array (ULA) comprising $M$ half-wavelength spaced antennas to serve a set of $K$ single-antenna user equipments (UEs). 
The direct UE-BS channel is ignored  because of poor propagation conditions or can be estimated and removed from the model via conventional CE methods by turning off RIS~\cite{8879620}. An RIS with $N = N_1 \, N_2$ elements, arranged in a half-wavelength-spaced uniform planar array (UPA) manner, is deployed to enable UEs connectivity.
It is assumed that both the BS and the UEs are in the far-field region of the RIS. 
In this setting, we focus on estimating the overall end-to-end UE-RIS-BS channel, when the UEs send unitary pilot towards the BS.
Following prior works~\cite{9844229,SCPD, PDMP,8879620}, we consider a quasi-static block-fading  scenario\footnote{{
While focused on the ideal model, the core principle of the proposed algorithm can be extended to non-ideal scenarios (e.g., Rayleigh fading and mobility), which is reserved for future work.}
}, where the propagation environment remains invariant during the coherence interval,  neglecting non-ideal effects such as mutual coupling, fabrication tolerances, and environment-dependent scattering.
An OFDM resource grid is allocated to the UEs for network operations. Among these resources, $P$ subcarriers and $Q$ time slots are reserved for CE, employed by the UEs to send orthogonal pilot sequences with no pilot contamination~\cite{massivemimobook}. For each of the $Q$ time slots, the RIS loads a different configuration to collect measurements under different environmental conditions.\footnote{Remark that the $Q$ time slots reserved for CE need to be consecutive and at the beginning of the overall resource grid. The CE can be acquired before performing RIS configuration optimization and resource allocation~\cite{Saggese2023framework}.}
Accordingly, the measurement matrix $\bm{Y}_{p}^k \in \mathbb{C}^{M\times Q}$ about UE $k$ and subcarrier $p$ can be expressed as~\cite{PDMP}:
\begin{align}
\begin{split}
  \bm{Y}_p^k &=\bm{G}_p \mathrm{diag}( \bm{h}_p^k ) \bm{\Theta} + \bm{W}_p^k \triangleq \bm{H}_p^k \bm{\Theta}  +\bm{W}_p^k, \,\,  \forall p,k. \label{eq:Yk}
      \end{split}
\end{align}
 where $\bm{G}_p \in\mathbb{C}^{M \times N}$ is the RIS-BS channel at the $p$-th subcarrier,
$\bm{h}_p^k\in\mathbb{C}^N$ is the UE-RIS channel at the $p$-th subcarrier of UE $k$; 
$\bm{H}_p^k=\bm{G}_p \mathrm{diag} (\bm{h}_p^k), \forall p$ is  the cascaded channel at the $p$-th subcarrier of UE $k$; 
$\bm{\Theta}  =[\bm{\theta}_1,...,\bm{\theta}_Q] \in \mathbb{C}^{N\times Q}$ is the configuration matrix of the RIS having elements $\theta_{n,q}=e^{j\varphi_{n,q}}$, with $\varphi_{n,q}$ representing the phase shift of the $n$-th RIS element at the ${q}$-th time slot;  
$\bm{W}_{p}^k$ is Additive White Gaussian Noise.
 \vspace{-0.3cm}
\subsection{Channel Model}
Considering the limited paths of mmWave system, $\bm{G}_p, \forall p$ and $\bm{h}_p^k, \forall p,k$ can be represented as~\cite{SCPD}
\begin{equation} \label{eq:Gfhf}
\begin{cases}
\bm{G}_p = \sum_{\ell = 1}^{L_1} {{\beta} _{{\ell}}}e^{ {- j\pi p \tau_{\ell}  }} {{\bm  a}_{{{M}}}}(\phi_{\ell}) {\bm a}_{{N_1,N_2}}\T (\omega_{\ell}, \psi_{\ell}), \\ 
\bm{h}_p^k = \sum_{{l=1}}^{{L^k_2}} {\beta _{l}^k}e^{ {- j\pi p\tau_{l,k}  }} {\bm a}_{N_1,N_2}(\omega_{l}^k, \psi_{l}^k). 
\end{cases}
\end{equation}
where ${{L_1}}$  denotes the number of RIS-BS path, while ${{L^k_2}}$  denotes the number of UE-RIS paths of UE $k$, respectively.
$\bm{a}_M(\cdot)$ and $\bm{a}_{N_1,N_2}(\cdot,\cdot)$ are the array steering vectors of the ULA and UPA, defined below in \eqref{eq:array1} and \eqref{eq:array}.
$\beta_{\ell}$ ($\kappa_{\ell}$) and 
$\beta _{{l}}^k$ ($\kappa_{l,k}$) are the complex channel gains (delays) of the $\ell$-th  RIS-BS path and the $l$-th UE-RIS path of UE $k$ with $\tau_{\ell} \triangleq {2 f_s \kappa_{\ell} /{P } }$, $\tau_{l}^k \triangleq {2 f_s \kappa_{l,k} /{P } }$, where $f_s$ is the sample frequency~\cite{SCPD}. 
$\phi_{\ell}$  and $\chi_{\ell}^a$ ($\chi_{\ell}^e$) are the cosine values of the angle of arrival (AoA) and 
the azimuth (elevation) angle of departure (AoD)  of the $\ell$-th RIS-BS  path  with $\omega_{\ell} \triangleq  \cos(\chi_{\ell}^a), \psi_{\ell} \triangleq  \sin ( \chi_{\ell}^a )  \cos (\chi_{\ell}^e)$. 
$\chi_{l,k}^{a}$ $(\chi_{l,k}^e)$ are the azimuth (elevation) AOA of the $l$-th  UE-RIS path of UE $k$ with $\omega_{l}^k \triangleq  \cos (\chi_{l,k}^a ), \psi_{l}^k \triangleq  \sin (\chi_{l,k}^a) \cos (\chi_{l,k}^e)$. 
${\bm a}_{{M}}(x_0)$ and ${\bm a}_{{N_1,N_2}}(x_1,x_2)$ are 
defined as
\begin{align}
    &{{\bm{a}}_{{{X}}}}( x_0 ) = {\left[ { 1,e^{ {- j\pi x_0}},\dots, e^{ {- j\pi \left( {X-1} \right) x_0}}   } \right]\T}/X, \label{eq:array1}
    \\
    &{{\bm{a}}_{{N_1,N_2}}}\left( x_1,x_2 \right) ={{\bm{a}}_{N_1}}\left( x_1 \right) \otimes {{\bm{a}}_{{{N_2}}}}\left( x_2 \right),
    \label{eq:array}
\end{align} 
where $X \in \{M,P,N_1,N_2\}$. 
The cascade channel ${\bm{H}}_p^k={\bm G}_p {\rm{diag}}({\bm h}_p^k)$ can be rewritten as--see~Appendix~A for details:
\vspace{-0.3cm}
\begin{equation}\label{eq:H1pk}
	\resizebox{0.89\hsize}{!}{$\displaystyle
	\begin{aligned}[b]
		{\bm{H}}_p^k & = \sum\nolimits_{\ell = 1}^{L_1} \sum\nolimits_{{l=1}}^{{L^k_2}}\beta_{\ell,l}^k  e^{ {- j\pi p \tau_{\ell,l}^k  }}  {{\bm{a}}_{{{M}}}}(\phi_{ \ell } ){\bm{a}}_{N_1,N_2}\T(\omega_{\ell,l}^k,\psi_{\ell,l}^k ), \\
		& = \sum\nolimits_{{u=1}}^{{U^k}} \beta_{u}^k  e^{ {- j\pi p \tau_{u}^k  }}  {{\bm{a}}_{{{M}}}}(\phi_{u} ){\bm{a}}_{N_1,N_2}\T(\omega_{u}^k,\psi_{u}^k ),
	\end{aligned}
	$}
\end{equation}
where $\{\phi_{\ell},\beta_{\ell,l}^k,\omega_{\ell,l}^k,\psi_{\ell,l}^k,\tau_{\ell,l}^k,\forall \ell,l,k\}$ are the parameters of the cascaded channel  and   $\{\phi_{u},\beta_{u}^k,\omega_{u}^k,\psi_{u}^k,\tau_{u}^k,\forall u,k\}$  are the the mapping parameters with  $u\triangleq(l-1)L_1+\ell, U^k = L_1L_2^k$, having the following mapping relationship~\cite{SCPD}
\begin{equation} \label{eq:para1}
\begin{cases}    
    &\beta_{\ell,l}^k\triangleq{\beta _{{\ell}}}  \beta _{l}^k \rightarrow  \beta_{u}^k,\quad \tau_{\ell,l}^k \triangleq\tau_{\ell}+\tau_{l}^k\rightarrow\tau_{u}^k,
    \\
    &\omega_{\ell,l}^k\triangleq \omega_{\ell}+\omega_{l}^k\rightarrow \omega_{u}^k,\quad \psi_{\ell,l}^k \triangleq\psi_{\ell}+\psi_{l}^k\rightarrow\psi_{u}^k,
    \\ 
    &\phi_{\ell,l}^k \triangleq \phi_{\ell}\rightarrow \phi_{u}, \quad  \forall \ell={\rm{mod}}({u},L_1),
    \end{cases} 
\end{equation} 
Here,  $\ell$ and $l$ index the  paths of the BS-RIS and RIS-UE links, respectively. Consequently, $\{\cdot\}_{\ell,l}^k$ represents the composite parameters formed by the specific pair of paths $(\ell, l)$. These tuple-indexed parameters are mapped to the linearized effective parameters indexed by $u$, as defined in eq.~\eqref{eq:para1}.

{Since $\bm{G}_p$ is physically shared by all UEs, its parameters 
serve as a common reference. Consequently, the derived structural features 
are \emph{mathematically invariant} across all users in this ideal model.
Based on this invariance observed in eq.~\eqref{eq:para1}, we classify the cascaded parameters into dual-structure features, specifically \emph{common} and \emph{offset} features:}

\emph{1) Common feature:}  There are only $L_1$ cascaded AoD parameters $\{ \phi_{l} , \forall \ell \} $ since $\phi_{\ell,l}^k = \phi_{ \ell }$. 
All the UEs  share the same cascaded AoD parameter, allowing 
us to combine all the UEs 
for the joint estimation of $\{ \phi_{l} , \forall \ell \} $.

\begin{figure}[!ht]
    \centering    
    \includegraphics[width=.95\columnwidth]{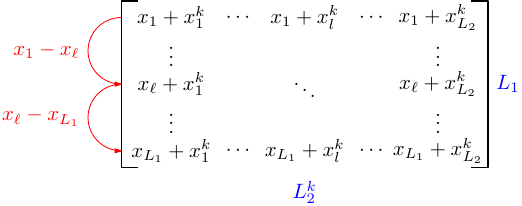}
    \caption{Offset feature of matrix  $\bm x^k, x \in \{ \tau,\omega,\psi\}$. }\label{fig:MURISCEAOA} 
    \vspace{-0.7cm}
\end{figure}

\emph{2) Offset feature:}   For $x \in \{ \tau,\omega,\psi\}$, define  $\bm x^k$ as the matrix form of $\{x_{\ell,l}^k \}$. 
As depicted in Fig.~\ref{fig:MURISCEAOA}, $\bm{x}^k$ exhibits invariant row differences, where the $\ell$-th and $\ell^\star$-th rows differ by a constant $x_{\ell}-x_{\ell^\star}$.
An analogous multiplicative structure holds for $\bm{\beta}^k$.
The offset feature can be used to estimate the number of channel paths and assist in obtaining other users' channel parameters, as detailed in later sections.

\subsection{Tensor Based System Model}
\begin{figure}[!h]\vspace{-0.3cm}
    \centering
    \includegraphics[width=0.499\textwidth]{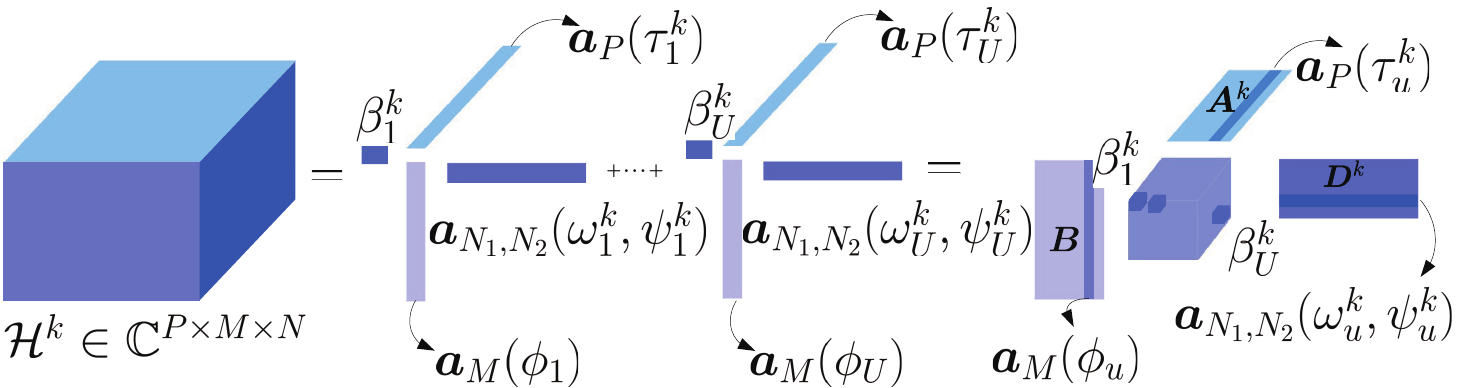}
    \caption{Tensor representation of the channel ${\cal H}^k$.}\label{fig:MUmeasuretensor} 
    \vspace{-0.55cm}
\end{figure}

For $P$ subcarriers, the channel $\{{\bm{H}}_p^k\}_{p=1}^P$ in  tensor form is:
\begin{align}
\label{eq:Htensor}
    {\cal  H}^k &=\sum\nolimits_{{u=1}}^{{U^k}} { \beta_{u}^k} {\bm a}_P(  {\tau_{u}^k} ) \circ {\bm a}_M({ \phi_{u}} )    \circ \bm  {\bm  a}_{N_1,N_2} ({\omega_{u}^k}, {\psi_{u}^k} ) 
    \nonumber
    \\
    &\triangleq \left[\kern-0.15em\left[ {\bm A}^k,{\bm B},{\bm  D}^k\mathrm{diag}(\text{vec}(\bm{\beta}^k)  ) \right]\kern-0.15em\right],
\end{align}
where $\bm{A}^k$, $\bm{B}$ and ${\bm  D}^k\mathrm{diag}(\text{vec}(\bm{\beta}^k)  )$ are the
 factor matrices with the following definition: 
 $\bm{A}^k = [\bm{a}_P ( \tau_{1}^k ), \dots, \bm{a}_P ( \tau_{U}^k)] \in \mathbb{C}^{P\times U}$, $\bm{B}  = [\bm{a}_M(\phi_{1}), \dots, \bm{a}_M(\phi_{U})] \in \mathbb{C}^{M\times U}$, $\bm  D^k  = [\bm  {\bm  a}_{N_1,N_2}(\omega_{1}^k,\psi_{1}^k),\dots,\bm  {\bm  a}_{N_1,N_2}(\omega_{U}^k,\psi_{U}^k)]$ $ \in \mathbb{C}^{N\times U}$.
 Note that the mapping $\phi_{\ell,l}^k \rightarrow \phi_{u}, \quad  \forall \ell={\rm{mod}}({u},L_1)$ of eq.~\eqref{eq:para1}, there are duplicate columns in matrix $\bm B$.
Fig.~\ref{fig:MUmeasuretensor} is the tensor representation of the channel ${\cal H}^k$.

Similarly, by substituting~\eqref{eq:H1pk} into~\eqref{eq:Yk}, the received signal can be represented by a tensor ${\cal \bm Y}^k \in \mathbb{C}^{P \times M \times Q}$ as~\cite{SCPD,tensorapp}
\begin{align}
\label{eq:Ytensor}
    {\cal{Y}}^k &=\sum\nolimits_{{u=1}}^{{U^k}} { \beta_{u}^k} {\bm a}_P({\tau_{u}^k})  \circ  {\bm a}_M({ \phi_{u}}) \circ \tilde{\bm{a}}_{N_1,N_2}({\omega_{u}^k}, {\psi_{u}^k }) +{\cal \bm W }\nonumber
    \\
    &\triangleq \left[\kern-0.15em\left[ { {\bm{A}^k},{{\bm{B}}},\bm{R}^k} \right]\kern-0.15em\right]+{\cal \bm W }^k={\cal{Z}}^k+{\cal{W}}^k,
\end{align}
where $\bm{R}^k \hspace{-0.2cm} \triangleq \hspace{-0.2cm}\bm{C}^k \mathrm{diag}(\text{vec}(\bm{\beta}^k)  )$ with $\hspace{-0.1cm}\bm{C}^k\hspace{-0.2cm}  = [\tilde{\bm{a}}_{N_1,N_2}(\omega_{1}^k,\psi_{1}^k) $, \dots ,$\tilde{\bm{a}}_{N_1,N_2}(\omega_{U}^k, \psi_{U}^k)]$, $\tilde{\bm{a}}_{N_1,N_2}(\omega_{u}^k,\psi_{u}^k)$$= \bm \Theta\T \bm{a}_{N_1,N_2}(\omega_{u}^k,\psi_{u}^k)$, and ${\cal{Z}}^k$ is the noiseless measurement of UE $k$.
Eq.~\eqref{eq:Ytensor} can be rewritten in the following unfolded and vectorized forms (detailed derivation in~ Appendix~B)
\begin{align}\label{eq:Y1m1}
\bm Y_{(1)}^k &= \bm A^k({{\bm{R}^k}} \odot {{\bm{B}}})\T+\bm W_{(1)}^k \in \mathbb{C}^{P \times MQ}, \\\label{eq:Y1m2}
\bm Y_{(2)}^k &= \bm B({{\bm{R}^k}} \odot {{\bm{A}^k}})\T+\bm W_{(2)}^k \in \mathbb{C}^{  M \times PQ}, \\\label{eq:Y1m3}
\bm Y_{(3)}^k &= \bm R^k({{\bm{B}}} \odot {{\bm{A}^k}})\T+\bm W_{(3)}^k\in \mathbb{C}^{ Q\times PM},
\\\label{eq:YMatrix2}
\text{vec}({\cal Y}^k)&=\bm G^k \text{vec}(\bm{\beta}^k) +  \text{vec}({\cal \bm W }^k) \in \mathbb{C}^{ PMQ},\vspace{-0.4cm}
\end{align}
where $\bm G^k=\left[\text{vec}({\cal G}_1^k), \dots, \text{vec}({\cal G}_U^k) \right]$ and ${\cal{G}}_u^k={\bm a}_P({\tau_{u,k}})  \circ  {\bm a}_M({ \phi_{u,k}}) \circ \tilde{\bm{a}}_{N_1,N_2}({\omega_{u,k}^a}, {\omega_{u,k}^e })$.
Eqs.~\eqref{eq:Y1m1},~\eqref{eq:Y1m2},~\eqref{eq:Y1m3} and~\eqref{eq:YMatrix2} are the  mode-$n$ unfoldings ($n=1, 2, 3$), and the vector form of tensor ${\cal \bm Y}^k$, respectively. They respectively express the BS, RIS, subcarrier and gain dimensions because they explicitly contain the cascaded AoD $\{\phi_{u}, \forall u,k\}$, AoA $\{\omega_{u}^k,\psi_{u}^k,\forall u,k\}$, delay $\{\tau_{u}^k,\forall u,k\}$ and gain parameters $\{{\beta}_{u}^k,\forall u,k\}$, respectively.  
This formalization enables the extraction of distinct parameters through strategic transformations between different dimensions, i.e, by using the so called MDT method.
\vspace{-0.25cm}
\section{Proposed  DS-MDT algorithm}
In this section, we present the proposed DS-MDT algorithm. 
Different from the direct tensor-based approach in~\cite{PDMP} that estimates the cascaded parameters independently, the proposed algorithm exploits the double-structure features to assist the CE process, through the following three steps: $i$) Combine all UEs' measurements to estimate the cascaded AoD parameter exploiting the common feature; $ii$) Estimate the remaining parameters of a reference UE and compute the offset feature;
$iii$) Estimate the channel parameters of all the other UEs based on the computed offset.
In the CE process, MDT is employed to explicitly reveal the corresponding dimensions; then,  high resolution MUSIC algorithm can be applied to obtain the channel parameters from the respective dimensions. \vspace{-0.4cm}
\subsection{Jointly estimate the cascaded AoD parameter $ \bm \phi$ }

We first focus on the mode-2 of $ {\cal Y}^k$, i.e., eq.~\eqref{eq:Y1m2},  to estimate $\bm \phi$, which is contained in  $\bm B$.
Recalling the duplicate columns of  $\bm B$ as analyzed in eq.~\eqref{eq:Htensor}, eq.~\eqref{eq:Y1m2} can be rewritten as
\begin{align}
\label{eq:Y14Y1m2}
    \bm Y_{(2)}^k &=\bm B  [\bm{\tilde f}_{1}^k, \dots, \bm{\tilde f}_{u}^k, \dots,\bm{\tilde f}_{U}^k]\T +\bm W_{(2)}^k \nonumber
    \\ 
    &=[\bm{a}_M(\phi_1), \dots, \bm{a}_M(\phi_{L_1})][\bm{f}_1^k,\dots,\bm{f}_{L_1}^k]\T +\bm W_{(2)}^k \nonumber 
    \\
    &=\sum\nolimits_{{\ell=1}}^{{L_1}}\bm{a}_M(\phi_\ell) (\bm{f}_\ell^k)\T \triangleq \bm {\Tilde{B}}(\bm F^k)\T+\bm W_{(2)}^k,
\end{align}
where   $\bm{f}_{\ell}^k=\sum_{{l=1}}^{L_2^K}{\bm{ \bar f}_{\ell,l}^k}$, 
${\bm{ \bar f}_{\ell,l}^k}$  and $\bm{\tilde f}_{u}^k$ have the  mapping   ${\bm{ \bar f}_{\ell,l}^k} \rightarrow  \bm{\tilde f}_{u}^k$ with $u\triangleq(\ell-1)L_1+l$. $\bm{\tilde f}_{u}^k$  is the $u$-th  columns of ${{\bm{R}^k}} \odot {{\bm{A}^k}}$. 

According to the common feature, all the UEs can be combined for the joint estimation of $\bm{\phi}^k$:
\begin{equation}\label{eq:Y14}
    \resizebox{0.88\hsize}{!}{$
    \bm Y_{(2)}=[\bm Y_{(2)}^1, \dots, \bm Y_{(2)}^K]= \bm {\Tilde{B}} [(\bm F^1)^{\mathrm{T}},\dots,(\bm F^K)^{\mathrm{T}}] +\bm W_{(2)}.
    $}
\end{equation}
From \eqref{eq:Y14}, 
$ \bm \phi$ can be estimated using the conventional MUSIC algorithm~\cite{C5Schmidt1986MultipleEL}.
{ This algorithm, however, requires prior knowledge of the number of paths $L_1$, which is typically unknown in practice. Therefore, the minimum description length (MDL) criterion is employed to estimate $L_1$~\cite{1164557}.}
Moreover, since $\bm Y_{(2)}$ has dimensions ${M \times PQK}$, the available number of samples ($PQK$) is  sufficient to estimate $\bm \phi$, leading to a  robust estimation even at low signal-to-noise ratios (SNRs).
\vspace{-0.4cm}

\subsection{Estimate  cascaded delay parameter $\bm \tau^k$ }\label{sec:esttau}

Recalling Fig.~\ref{fig:MURISCEAOA}, $\bm{\tau}^k$ possesses the offset feature.
{To leverage this structure, we designate a reference UE to assist parameter extraction. 
For robustness, the reference index $k_{\text{ref}}$ is determined per coherence block by maximizing the received signal energy prior to CE:
{
\setlength{\abovedisplayskip}{3pt}
\setlength{\belowdisplayskip}{3pt}
\begin{equation}
    k_{\text{ref}} = \arg \max_{k} \| \mathcal{Y}^k \|_F^2.
\end{equation}
}
Without loss of generality, we sort the user indices such that the reference UE corresponds to $k=1$. 
This selection identifies the user with the highest instantaneous SNR to serve as a reliable anchor for UEs with weaker channel conditions, thereby enhancing overall estimation performance, while the block-wise update schedule inherently prevents switching instability. 
Consequently, for non-reference UEs ($k > 1$), only the first row requires explicit computation, while the remaining rows are derived via the offset feature, significantly reducing the computational complexity.}

We first consider the reference UE ($k=1$). 
Assume $\hat{\bm B}$ is the estimator of $\bm{\Tilde{B}} $ obtained from estimated $\bm \phi$. By left multiplying $(\hat{\bm{B}})^\dagger $ to eq.~\eqref{eq:Y14Y1m2} and considering $k=1$, we obtain
\begin{equation}
\label{eq:Y111}
    \bm{\tilde Y}_{(2)}^1 \approx (\bm F^1)\T +\bm {\tilde W}_{(2)}^1,
\end{equation}
where $\bm {\tilde Y}_{(2)}^1=(\hat{\bm{B}})^\dagger\bm Y_{(2)}^1$ and $\bm {\tilde W}_{(2)}^1=(\hat{\bm{B}})^\dagger\bm W_{(2)}^1$.
We estimate $\bm \tau^1$  based on the known estimation $\hat{\bm{\phi}}$ rather than estimating them  independently as in~\cite{PDMP}.
Recalling the definition of $\bm F^k$ in eq.~\eqref{eq:Y14Y1m2}, $\bm F^1$ separates $U^1$ delay parameters into its $L_1$ columns, $\bm{f}_{\ell}^1=\sum_{{l=1}}^{L_2^1}{\bm{ \bar f}_{\ell,l}^1}$ $\forall \ell$, each containing  $L_2^1$ delay parameters to be estimated. 
Thus, each column of  $\bm {\tilde Y}_{(2)}^1$ can be converted into a matrix through\vspace{-3mm}
\begin{align}
\label{eq:s5YMatrixl}
    &\text{Mat}\big(\bm {\tilde Y}_{(2)}^1(:,l)\big) \approx \text{Mat}\big(\sum\nolimits_{{l=1}}^{L_2^1}{\bm{ \bar f}_{\ell,l}^1}\big) \nonumber
    \\ 
   & = [\bm{a}_P(\tau_{\ell,1}^1), \dots, \bm{a}_P(\tau_{\ell,L_2^1}^1)][\bm{r}_{\ell,l}^1, \dots, \bm{r}_{\ell,l}^1]\T
    \triangleq \bm A_l^1 \bm R_l^1.
\end{align}

\vspace{-3mm}

Finally, $\bm \tau^1$ can be obtained using the MUSIC algorithm on eq.~\eqref{eq:s5YMatrixl}. 
Since the limited samples in eq.~\eqref{eq:s5YMatrixl}  render MDL ineffective for determining $L^1_2$,
we adopt an offset-based approach.
We set $\hat L_2^1$ slightly larger than $L_2^1$ (justified in~\cite{Mo2025MultidimensionalPE})  and  compute the relative offsets.
{
While the computed offsets theoretically maintain the consistency described in Section II,
practical observations suffer from noise-induced deviations.
To address this, we employ median absolute deviation (MAD) filtering (via the MATLAB function \texttt{rmoutliers}). This step isolates the valid estimators by filtering out outliers; the retained values correspond to $\bm{\tau}^1$ and their count yields $L_2^1$.
}
For non-reference UEs, only the first row of  $\bm \tau^k$ needs to be calculated,
while the remaining rows can be generated exploiting the offset feature.
Although some UEs may have low SNR, the reference UE assisting other UEs can improve estimation accuracy. 

\vspace{-0.4cm}

\subsection{Estimate cascaded AoA parameters 
 $\{\bm{\omega}^{k}, \bm{\psi}^{k}\}$ }
We then turn to RIS dimension~\eqref{eq:Y1m3}, i.e., the mode-3 matrix of $ {\cal \bm Y}^k $, to estimate $\bm{\omega}^{k}$ and $ \bm{\psi}^{k}$. 
With the same idea of estimating $\bm \tau^k$, we first consider the reference UE ($k=1$).

After $\hat {\bm{A}}^1$ and  $ \hat {\bm{B}} $ are obtained from estimated $\hat{\bm{\tau}}^1$ and $\hat{\bm{\phi}}$, respectively, the 
least squares (LS) can be used on eq.~\eqref{eq:Y1m3} to separate different paths of $(\bm{\omega}^{1}, \bm{\psi}^{1})$, i.e.,~\cite{PDMP}: 
    \begin{align} \label{eq:s5LS}
        \hat{\bm{R}}^1= \bm{Y}_{(3)}^1\big[(\hat {\bm{B}} \odot \hat {\bm{A}}^1)^{T}\big]^{\dagger}.
    \end{align}
The $u$-th column of $\hat{\bm{R}}^1$ exhibits a single-path structure $\hat{\bm{r}}^1_u\triangleq\beta_{u}^1\tilde{\bm{a}}_{N_1,N_2}(\omega_{u}^1, \psi_{u}^1)$.  The associated parameters $({\omega}_{u}^1, {\psi}_{u}^1)$ are estimated via~\cite{SCPD}:
\begin{equation}\label{eq:s5crror}
	\resizebox{0.89\hsize}{!}{$\displaystyle
	(\hat{\omega}_{u}^1, \hat{\psi}_{u}^1)=
\arg \max _{\omega_{u}^1,\psi_{u}^1} {|\tilde{\bm{a}}_{N_1,N_2}^\mathrm{H}(\omega_{u}^1,\psi_{u}^1) \hat{\bm{r}}^1_u|}/{\left\|\tilde{\bm{a}}_{N_1,N_2}^\mathrm{H}(\omega_{u}^1,\psi_{u}^1)\right\|_2}.$}
\end{equation}
For non-reference UEs, $\{\bm{\omega}^{k}, \bm{\psi}^{k}\}$ can be obtained by offset feature with the same  method estimating $\bm \tau^k, 2 \leq k \leq K$.

\vspace{-0.4cm}
\subsection{Estimate $\bm \beta^k$  }
\label{sec:beta}
We turn to the gain dimension~\eqref{eq:YMatrix2} to  estimate $\bm \beta^k$.
Unlike other parameters, the components in ${\bm{\beta}}^k$ are coupled through $\text{vec}( {\cal \bm Y}^k)$ in eq. (12), which prevents us from estimating them in groups. 
Therefore, 
${\bm{\beta}}^k$ is estimated separately for each UE, by LS. Specifically,
after $\hat {\bm{A}}^k$,  $ \hat {\bm{B}} $ and  $ \hat {\bm{D}}^k $ are obtained from the estimated ${\bm{\tau}}^k$, ${\bm{\phi}}$ and $\{\bm{\omega}^{k}, \bm{\psi}^{k}\}$, $\bm \beta^k$ is obtained as 
    \begin{align} \label{eq:s5LS1}
        \text{vec}(\hat{\bm{\beta}}^k)= \big[ \bm G^k \big]^{\dagger} \text{vec}({\cal Y}^k). 
    \end{align}
Similarly to $L_2^1$, $L_2^k$ is obtained from $\hat{{\bm{\beta}}}^k$ via the offset feature.

\begin{algorithm}[b!]
\footnotesize
\caption{DS-MDT Algorithm}\label{alg:DS-MDT}
\KwInput{${\cal Y}^k(1 \leq k \leq K), {\bm \Theta}$, $\hat L_2^k$}
\SetKwInput{KwInitialize}{Initialize}
Compute and set  the UE with max received power as reference UE\;
Estimate $\bm \phi$  by MUSIC  via~\eqref{eq:Y14} after obtaining $ L_1$ via MDL\;
Divide $\bm{\tau}^1$ into $ L_1$ groups via~\eqref{eq:Y111}\;
Estimate $\bm{\tau}^1$ by MUSIC   via~\eqref{eq:s5YMatrixl}\;
Get offset value  and $ L_2^1$  from estimated $\bm{\tau}^1$\;
Estimate the first row of $\bm{\tau}^k(2 \leq k \leq K)$ by MUSIC via~\eqref{eq:s5YMatrixl}\;
Estimate $\bm{\tau}^k(2 \leq k \leq K)$ from the estimated first row of $\bm{\tau}^k$ and offset feature\;
Divide  $\bm{\omega}^{1}, \bm{\psi}^{1}$ into $U^1$ groups via~\eqref{eq:s5LS}\;
Estimate $\bm{\omega}^{1}, \bm{\psi}^{1}$  via~\eqref{eq:s5crror} \;
Get offset feature of  $\bm{\omega}^{k}, \bm{\psi}^{k}$ from estimated $\bm{\omega}^{1}, \bm{\psi}^{1}$\;
Estimate the first row of  $\bm{\omega}^{k},\bm{\psi}^{k}(2 \leq k \leq K)$  by MUSIC  via~\eqref{eq:s5crror}\;
Estimate $\bm{\omega}^{k}, \bm{\psi}^{k}(2 \leq k \leq K)$  from the estimated first row of $\bm{\omega}^{k}, \bm{\psi}^{k}(2 \leq k \leq K)$ and offset feature\;
Estimate $\bm{\beta}^{k}(1 \leq k \leq K)$ via~\eqref{eq:s5LS1} and obtain $ L_2^k (2 \leq k \leq K)$ using offset feature\;
Repeat 13 after removing the outliers if estimated $ L_2^k  \neq$ initial \;
{Compute  $\mathbb{I}_{\text{sys}}$ via~\eqref{eq:indicator}.  Switch to alternative  if anomalous}\;
\KwOutput{ ${\cal {\hat H}}^k(1 \leq k \leq K)$}
\normalsize  \vspace{-0.05cm}
\end{algorithm}

{To ensure robustness against severe outliers, we define the structural validity criterion. 
Let $\mathcal{T}(\bm{x}) = \ln|\bm{x}|$ (element-wise) for $\bm{x} = \bm{\beta}^k$, and $\mathcal{T}(\bm{x}) = \bm{x}$ for $\bm{x} = \bm{\tau}^k$. 
The structural inconsistency is quantified by $\mathcal{D}(\bm{x}) = \| \mathbf{d} - \bar{\mu} \|_\infty$, where $\mathbf{d} = [\mathcal{T}(\bm{x})]_{1,:} - [\mathcal{T}(\bm{x})]_{2,:}$ denotes the difference between the first two rows, and $\bar{\mu}$ denotes the  mean of $\mathbf{d}$.
A user is deemed reliable only if both offset and multiplicative structures are consistent. 
Consequently, the system validity indicator $\mathbb{I}_{\text{sys}}$ follows a robust $k_0$-out-of-$K$ voting logic:
\begin{equation}\label{eq:indicator}
    \mathbb{I}_{\text{sys}} = \mathbb{I} \Big\{ \sum\nolimits_{k=1}^{K} \Big( \prod\nolimits_{\bm{x} \in \mathcal{S}_k} \mathbb{I} \{ \mathcal{D}(\bm{x}) < \epsilon \} \Big) \ge k_0 \Big\},
\end{equation}
where $\mathcal{S}_k$ is the feature set of the $k$-th user and $\epsilon$ is the tolerance threshold.
By imposing joint constraints on both additive and multiplicative structures, the indicator $\mathbb{I}_{\text{sys}}$ effectively filters unstructured noise. Supported by the $k_0$-out-of-$K$ voting logic,  this mechanism achieves a near 100\% anomaly rejection rate even with a relaxed threshold (e.g., $\epsilon=0.1$ and $k_0=0.5K$).
}
Finally,  ${\cal{H}}^k$ can be obtained by estimating
$\{ \bm{\phi}, \bm{\tau}^k,  \bm{\omega}^{k}, \bm{\psi}^{k},\bm \beta^k \}$, as
summarized in \textbf{Algorithm~\ref{alg:DS-MDT}}.

\section{Simulation Results}\label{Sec:simulation results}  
We  now verify  the performance of the proposed DS-MDT algorithm.
The simulation parameters are as follows, set similar to~\cite{SCPD}:
$f_c=28\,\mathrm{GHz}, M=64, N=16\times 16, K=8, L_1=L_2^k=3$.
We set $d_\ell = 30$ m, $d_l \sim \mathcal{U}[20, 40]$ m, and gains $\beta_{\ell}, \beta_{l}^k \sim \mathcal{CN}\big(0, (c/4 \pi d f_c)^2\big)$ with $d \in \{d_\ell, d_l\}$.
Distributions for $\varphi _{n,q}$ and $\{\bm{\tau}^k, \bm{\phi}, \bm{\omega}^{k}, \bm{\psi}^{k}\}$ are $\mathcal{U}[0, 2\pi)$ and $\mathcal{U}[0, 1)$, respectively.
We use $P=128$, $Q=16$, $\mathrm{SNR}=10\,\mathrm{dB}$, and $\hat{L}_2^k=4 > L_2^k$ by default.
The proposed scheme is compared with a DS-MDT with known path number (DS-MDT-KPN) baseline, the state-of-the-art methods SCPD~\cite{SCPD} and PDMP~\cite{PDMP}, and two heuristics: 
$i$) PMDP-TT: as PDMP~\cite{PDMP}, except that $\bm  \tau^k$ are estimated  based on the known estimation $\hat{\bm{\phi}}^k$; $ii$) MTensor: the proposed approach is applied to estimate $\bm \phi$ and $\bm \tau^k$, while the other parameters are obtained through PDMP~\cite{PDMP}.

Prior to benchmarking against the baselines, Fig.~\ref{fig:paths} validates the path estimation accuracy and the  reference UE mis-selection robustness.
Fig.~\ref{fig:epsFigpath} presents the path estimation success rate (PESR) versus SNR. 
{While the PESR of $L_1$ approaches $100\%$ due to the multi-user array gain, 
$L_2^1$ and $L_2^k$  exhibit lower performance as noise submerges weak multipath components,  preventing the extraction of delay parameters required for offset consistency.
Moreover, the PESR of $L_2^1$ (from additive $\bm{\tau}^1$) is higher than that of $L_2^k$ (from multiplicative ${\bm{\beta}}^k$), as the multiplicative form is more sensitive to estimation error.
Fig.~\ref{fig:epsFigpath1} presents the estimated number of channel paths (ENCP) statistics, where vertical lines and error bars denote the mean and standard deviation, respectively. 
The average ENCP is about 3, matching the actual length with low variance.  
This minimal variance confirms that the distribution of offset errors remains highly concentrated.
Despite sporadic path estimation errors, their impact is limited due to the low magnitude of the corresponding channel gains, corroborated by the good CE performance shown next.
Fig.~\ref{fig:epsRatio} assesses robustness against unexpected signal fluctuations, where $P_{\text{mis}}$ denotes the probability of selecting the worst-case UE. While NMSE  degrades with increasing $P_{\text{mis}}$, the algorithm maintains reliable estimation without catastrophic failure even at $P_{\text{mis}}=100\%$, confirming its robustness.
}

Fig.~\ref{fig:NMSE} shows the performance comparison in terms of NMSE.
Fig.~\ref{fig:epsFigP} presents the NMSE as a function of the available subcarriers $P$. Performance improves as $P$ increases, for all algorithms except SCPD. The latter depends on the estimation of factor matrices--whose dimension grows with 
$P$--leading to a larger number of unknowns and, thus, degraded performance. Conversely, the probability of angular ambiguity in the PDMP and PDMP-TT algorithms decreases, when $P$ increases.  Also, PDMP-TT outperforms PDMP, due to the grouping strategy of Sec.~\ref{sec:esttau} reducing the number of parameters estimated from $U^1$ to $L_2^1$, and thereby lowering the potential for angular ambiguity.
For the DS-MDT and MTensor algorithms, increasing $P$ augments the number of observations, thereby improving estimation precision. 
Moreover, DS-MDT outperforms MTensor, benefiting from the grouping operation and from leveraging the UE with the highest SNR to assist other UEs.
Fig.~\ref{fig:epsFigSNR} shows the NMSE as a function of the SNR.
The performance of SCPD improves rapidly with the increasing SNR, because its estimation process is strongly affected by noise since the delay and angle parameters are derived from the estimated factor matrices.
The performance of PDMP and PDMP-TT is governed primarily by the angular ambiguity phenomena, affected primarily by $P$ and almost unaffected by SNR and $Q$~\cite{PDMP}.
DS-MDT demonstrates superior performance by exploiting both the multi-dimensional structure of tensor and the dual-structure features among parameters of different UEs.
Fig.~\ref{fig:epsFigQ} evaluates the impact of pilot overhead $Q$
showing that DS-MDT exhibits the best performance even with short pilot durations.
For all tests, {DS-MDT-KPN} attains the best performance, followed closely by {DS-MDT}, proving that the proposed solution achieve remarkable performance even with an unknown number of paths.

\begin{figure}\vspace{-0.4cm}
    \centering
        \subfloat[c][SNR vs. PESR.\label{fig:epsFigpath}]{\definecolor{mycolor1}{rgb}{1.00000,0.00000,1.00000}%
\definecolor{mycolor2}{rgb}{0.00000,1.00000,1.00000}%
\definecolor{mycolor3}{rgb}{0.12941,0.12941,0.12941}%
\definecolor{mycolor4}{rgb}{1.00000,1.00000,0.00000}%

\begin{tikzpicture}

\begin{axis}[%
width=\fwidthq,
scale only axis,
xmin=0,
xmax=20,
xlabel={SNR [dB]},
ymin=75,
ymax=100,
ylabel={PESR [\%]},
xmajorgrids,
ymajorgrids,
legend style={
    at={(0.98,0.02)},      
    anchor=south east,
    legend columns=2,      
    font=\fontsize{3}{4}\selectfont, 
    legend image post style={scale=0.43},     
    /tikz/every even column/.append style={column sep=1ex},
    label style={font=\fontsize{3}{4}\selectfont},          
    inner sep=1.5pt,      
    row sep=-1pt,         
    draw=black!40,
    fill=white,
    fill opacity=0.85,
    align=left
},
ylabel style={inner sep=3pt, yshift=-0.6em},
ylabel shift=-3pt,
]


\addlegendimage{draw=none}
\addlegendentry{}

\addlegendimage{color=red, mark=square, mark options={solid, red, mark size=3pt}}
\addlegendentry{$L_1$}

\addlegendimage{color=mycolor1, dashed, mark=star, mark options={solid, mycolor1, mark size=3pt}}
\addlegendentry{$L_2^1$}

\addlegendimage{color=green, dashdotted, mark=x, mark options={solid, green, mark size=3pt}}
\addlegendentry{$L_2^{2}$}

\addlegendimage{color=blue, dotted, mark=triangle, mark options={solid, blue, mark size=3pt}}
\addlegendentry{$L_2^{3}$}

\addlegendimage{color=mycolor2, dashed, mark=diamond, mark options={solid, mycolor2, mark size=3pt}}
\addlegendentry{$L_2^{4}$}

\addlegendimage{color=mycolor1, dashdotted, mark=square, mark options={solid, mycolor1, mark size=3pt}}
\addlegendentry{$L_2^{5}$}

\addlegendimage{color=mycolor3, dotted, mark=o, mark options={solid, mycolor3, mark size=3pt}}
\addlegendentry{$L_2^{6}$}

\addlegendimage{color=mycolor4, dashed, mark=triangle, mark options={solid, rotate=180, mycolor4, mark size=3pt}}
\addlegendentry{$L_2^{7}$}

\addlegendimage{color=red, mark=x, mark options={solid, red, mark size=3pt}}
\addlegendentry{$L_2^{8}$}



\addplot [color=red, mark=square, mark options={solid, red}, forget plot]
  table[row sep=crcr]{0 96.75\\ 5 98.52\\ 10 99.33\\ 15 99.52\\ 20 99.77\\};

\addplot [color=mycolor1, dashed, mark=star, mark options={solid, mycolor1}, forget plot]
  table[row sep=crcr]{0 93.98\\ 5 96.38\\ 10 96.86\\ 15 98.04\\ 20 98.19\\};

\addplot [color=green, dashdotted, mark=x, mark options={solid, green}, forget plot]
  table[row sep=crcr]{0 89.96\\ 5 92.47\\ 10 94.20\\ 15 96.16\\ 20 96.97\\};

\addplot [color=blue, dotted, mark=triangle, mark options={solid, blue}, forget plot]
  table[row sep=crcr]{0 90.55\\ 5 93.13\\ 10 94.87\\ 15 96.90\\ 20 96.93\\};

\addplot [color=mycolor2, dashed, mark=diamond, mark options={solid, mycolor2}, forget plot]
  table[row sep=crcr]{0 89.96\\ 5 92.99\\ 10 94.31\\ 15 96.09\\ 20 96.86\\};

\addplot [color=mycolor1, dashdotted, mark=square, mark options={solid, mycolor1}, forget plot]
  table[row sep=crcr]{0 89.92\\ 5 92.62\\ 10 94.50\\ 15 96.90\\ 20 97.30\\};

\addplot [color=mycolor3, dotted, mark=o, mark options={solid, mycolor3}, forget plot]
  table[row sep=crcr]{0 89.26\\ 5 92.51\\ 10 94.13\\ 15 96.09\\ 20 96.86\\};

\addplot [color=mycolor4, dashed, mark=triangle, mark options={solid, rotate=180, mycolor4}, forget plot]
  table[row sep=crcr]{0 88.86\\ 5 92.36\\ 10 94.17\\ 15 96.12\\ 20 96.79\\};

\addplot [color=red, mark=x, mark options={solid, red}, forget plot]
  table[row sep=crcr]{0 87.90\\ 5 91.92\\ 10 94.13\\ 15 96.31\\ 20 96.97\\};

\end{axis}
\end{tikzpicture}} 
    \subfloat[c][SNR vs. ENCP.\label{fig:epsFigpath1}]{\definecolor{mycolor3}{rgb}{0.20000,0.80000,0.40000}%

\begin{tikzpicture}

\begin{axis}[%
width=\fwidth,
scale only axis,
ybar,
bar width=3pt,
xmin=0.5, xmax=4.5,
xtick={1,2,3,4},
xticklabels={{0},{5},{10},{15}},
xlabel={SNR [dB]},
ymin=0, ymax=3.5,
ylabel={ENCP},
tick align=inside, 
ymajorgrids,       
legend style={
    at={(0.5,0)}, 
    anchor=south, 
    legend columns=2, 
    /tikz/every even column/.append style={column sep=3mm},
    nodes={inner sep=1pt}
},
ylabel style={inner sep=0pt, xshift=-0.em},
legend image code/.code={\draw [#1] (0cm,-0.1cm) rectangle (0.15cm,4pt);},
error bars/y dir=both,
error bars/y explicit,
error bars/error bar style={line width=0.5pt, black},
error bars/error mark options={
    rotate=90,
    mark size=1.0pt,
    line width=0.5pt,
    black
},
ylabel shift=-3pt,
]

\addplot[fill=cyan, draw=black] 
table[row sep=crcr, y error plus index=2, y error minus index=3] {%
1	2.96753965326448	0.177251972636195	0.177251972636195\\
2	2.9852452969384	    0.120591732745121	0.120591732745121\\
3	2.99336038362228	0.0812278625083492	0.0812278625083492\\
4	2.99520472150498	0.0690944627991817	0.0690944627991817\\
};
\addlegendentry{$L_1$}

\addplot[fill=red!20!brown, draw=black] 
table[row sep=crcr, y error plus index=2, y error minus index=3] {%
1	3.06012541497602	0.237762911273935	0.237762911273935\\
2	3.03614902250092	0.186695280067805	0.186695280067805\\
3	3.0313537440059	    0.174304026454323	0.174304026454323\\
4	3.01954998155662	0.138473292584616	0.138473292584616\\
};
\addlegendentry{$L_2^1$}

\addplot[fill=mycolor3, draw=black] 
table[row sep=crcr, y error plus index=2, y error minus index=3] {%
1	3.10296674922274	0.30737174727227	0.30737174727227\\
2	3.07213995889761	0.262768412143825	0.262768412143825\\
3	3.05633134847447	0.231251200771755	0.231251200771755\\
4	3.03599093639669	0.187119229790294	0.187119229790294\\
};
\addlegendentry{$[L_2^k]_{k>1}$}

\end{axis}
\end{tikzpicture}} 
    \subfloat[c][SNR vs. NMSE ($P_{\text{mis}}$).\label{fig:epsRatio}]{\definecolor{mycolor1}{rgb}{1.00000,0.00000,1.00000} 
\definecolor{mycolor2}{rgb}{0.00000,1.00000,1.00000} 

\begin{tikzpicture}

\begin{axis}[
width=\fwidthq,
scale only axis,
xmin=0, xmax=20,
ymin=-22, ymax=-5,
xlabel={SNR [dB]},
ylabel={NMSE [dB]},
xmajorgrids,
ymajorgrids,
label style={font=\footnotesize},      
tick label style={font=\tiny}, 
legend style={
    at={(0.98,0.98)},      
    anchor=north east,   
    legend columns=1,   
    font=\fontsize{3}{4}\selectfont, 
    legend image post style={scale=0.5}, 
    label style={font=\fontsize{3}{4}\selectfont},
    inner sep=1.5pt,
    row sep=-1pt,   
    draw=black!40,
    fill=white,
    fill opacity=0.85,
    align=left
},
ylabel style={inner sep=1pt, yshift=-0.5em}
]

%

\addlegendimage{color=red, mark=square, mark options={solid, red, mark size=3pt}}
\addlegendentry{$0\%$}

\addlegendimage{color=mycolor1, dashed, mark=star, mark options={solid, mycolor1, mark size=3pt}}
\addlegendentry{$20\%$}

\addlegendimage{color=green, dashdotted, mark=x, mark options={solid, green, mark size=3pt}}
\addlegendentry{$40\%$}

\addlegendimage{color=blue, dotted, mark=triangle, mark options={solid, blue, mark size=3pt}}
\addlegendentry{$60\%$}

\addlegendimage{color=mycolor2, dashed, mark=diamond, mark options={solid, mycolor2, mark size=3pt}}
\addlegendentry{$80\%$}

\addlegendimage{color=mycolor1, dashdotted, mark=square, mark options={solid, mycolor1, mark size=3pt}}
\addlegendentry{$100\%$}

\addplot [color=red, mark=square, mark options={solid, red}, mark size=0.6pt, forget plot]
 table[row sep=crcr]{
0	-15.85821872\\
5	-17.02103873\\
10	-18.56017534\\
15	-19.53975474\\
20	-21.36283640\\
};

\addplot [color=mycolor1, dashed, mark=star, mark options={solid, mycolor1}, mark size=0.6pt, forget plot]
 table[row sep=crcr]{
0	-14.93628448\\
5	-16.40220730\\
10	-17.96355025\\
15	-19.12873247\\
20	-20.81442649\\
};

\addplot [color=green, dashdotted, mark=x, mark options={solid, green}, mark size=0.6pt, forget plot]
 table[row sep=crcr]{
0	-14.17620295\\
5	-15.86065643\\
10	-17.43907606\\
15	-18.75326882\\
20	-20.32756773\\
};

\addplot [color=blue, dotted, mark=triangle, mark options={solid, blue}, mark size=0.6pt, forget plot]
 table[row sep=crcr]{
0	-13.52954279\\
5	-15.37920958\\
10	-16.97116970\\
15	-18.40769817\\
20	-19.88982737\\
};

\addplot [color=mycolor2, dashed, mark=diamond, mark options={solid, mycolor2}, mark size=0.6pt, forget plot]
 table[row sep=crcr]{
0	-12.96680808\\
5	-14.94584827\\
10	-16.54880808\\
15	-18.08760961\\
20	-19.49219643\\
};

\addplot [color=mycolor1, dashdotted, mark=square, mark options={solid, mycolor1}, mark size=0.6pt, forget plot]
 table[row sep=crcr]{
0	-12.46869594\\
5	-14.55183345\\
10	-16.16390600\\
15	-17.78950178\\
20	-19.12793762\\
};

\end{axis}
\end{tikzpicture}} 
     \caption{Path estimation accuracy and reference UE mis-selection robustness.}
     \label{fig:paths}
    \vspace{-.4cm}
\end{figure}
\begin{figure}[t]
    \centering
    \subfloat{\begin{tikzpicture}
\begin{axis}[
    width=0cm,
    height=0cm,
    axis line style={draw=none},
    tick style={draw=none},
    at={(0,0)},
    scale only axis,
    xmin=0,
    xmax=1,
    xtick={},
    ymin=0,
    ymax=1,
    ytick={},
    axis background/.style={fill=white},
    legend style={at={(0, 0)}, anchor=center, draw=none, /tikz/every even column/.append style={column sep=1mm}, legend columns=3},
    ]

\addlegendimage{draw=black, mark=triangle}
\addlegendentry{DS-MDT}
\addlegendimage{draw=pink, dashed, mark=star}
\addlegendentry{PDMP}
\addlegendimage{color=green, dashed, mark=x, mark options={solid, green}}
\addlegendentry{SCPD}
\addlegendimage{color=red, mark=star}
\addlegendentry{PDMP-TT}
\addlegendimage{color=magenta, dashed, mark=square, mark options={solid, magenta}}
\addlegendentry{MTensor}
\addlegendimage{color=indigo, dashed}
\addlegendentry{DS-MDT-KPN}

\end{axis}
\end{tikzpicture}}\vspace{-.5cm}\\
    \setcounter{subfigure}{0}
    \subfloat[c][$P$ vs. NMSE.\label{fig:epsFigP}]{\begin{tikzpicture}

\begin{axis}[%
width=\fwidth,
scale only axis,
xmin=64,
xmax=256,
xlabel={$P$},
ymin=-25,
ymax=-5,
ylabel={NMSE [dB]},
xmajorgrids,
ymajorgrids,
legend style={legend cell align=left, align=left},
ylabel style={inner sep=0pt, yshift=-0.2em}
]
\addplot [color=black, mark=triangle]
  table[row sep=crcr]{%
32	-9.18911534224756\\
64	-14.2303962438594\\
96	-16.6429140382544\\
128	-19.1673124958725\\
160	-19.8915513385484\\
192	-20.3151705144606\\
224	-20.468366495785\\
256	-22.103524092077\\
};

\addplot [color=pink, dashed, mark=star]
  table[row sep=crcr]{%
32	-5.51918583925433\\
64	-7.52740910999694\\
96	-8.67912467802516\\
128	-9.54397609662519\\
160	-10.1739216834189\\
192	-10.5774223772839\\
224	-10.8465264546487\\
256	-11.4253737169727\\
};

\addplot [color=green, dashed, mark=x, mark options={solid, green}]
  table[row sep=crcr]{%
32	-11.4558341524118\\
64	-11.0546038009644\\
96	-10.0841007573161\\
128	-9.33727344985761\\
160	-8.49465318519723\\
192	-8.04486354391754\\
224	-7.53776826873326\\
256	-7.1043588136418\\
};

\addplot [color=red, mark=star]
  table[row sep=crcr]{%
32	-8.35385738382262\\
64	-10.1985725717842\\
96	-11.0302050806397\\
128	-11.3431585250212\\
160	-11.6576910791944\\
192	-12.0049272827524\\
224	-11.9737017052969\\
256	-12.343322570736\\
};

\addplot [color=magenta, dashed, mark=square, mark options={solid, magenta}]
  table[row sep=crcr]{%
32	-10.690514311214\\
64	-13.6573027258688\\
96	-15.4006154097984\\
128	-16.7716009084985\\
160	-17.8154654322666\\
192	-18.573669667782\\
224	-19.6801374853918\\
256	-19.893036488314\\
};

\addplot [color=indigo, dashed]
  table[row sep=crcr]{%
32	-12.677726620853\\
64	-18.2084197218255\\
96	-19.9239758426526\\
128	-21.3128884659419\\
160	-21.5295722999473\\
192	-22.2346880681769\\
224	-22.5319790499487\\
256	-23.6625060941765\\
};

\end{axis}
\end{tikzpicture}%
}
    \subfloat[c][SNR vs. NMSE.\label{fig:epsFigSNR}]{\begin{tikzpicture}
\begin{axis}[%
width=\fwidth,
scale only axis,
xmin=0,
xmax=20,
xlabel={SNR [dB]},
ymin=-25,
ymax=-5,
xmajorgrids,
ymajorgrids,
yticklabels={},
]
\addplot [color=black, mark=triangle]
  table[row sep=crcr]{%
0	-15.8582187172732\\
5	-17.0210387301406\\
10	-18.5601753388946\\
15	-19.5397547375952\\
20	-21.3628363988616\\
};

\addplot [color=pink, dashed, mark=star]
  table[row sep=crcr]{%
0	-9.47520577447354\\
5	-9.5220775487517\\
10	-9.4996168515248\\
15	-9.49324935002642\\
20	-9.47907323260805\\
};

\addplot [color=green, dashed, mark=x, mark options={solid, green}]
  table[row sep=crcr]{%
0	-3.31071545427716\\
5	-5.80467457289674\\
10	-9.20260482745507\\
15	-13.1315818685946\\
20	-16.6824072855719\\
};

\addplot [color=red, mark=star]
  table[row sep=crcr]{%
0	-11.2861981521246\\
5	-11.2888984239131\\
10	-11.3388344766812\\
15	-11.2760887154798\\
20	-11.3359698569638\\
};

\addplot [color=magenta, dashed, mark=square, mark options={solid, magenta}]
  table[row sep=crcr]{%
0	-15.052811155592\\
5	-15.7245695852825\\
10	-16.3869323673669\\
15	-16.8412980603602\\
20	-17.1468335459399\\
};

\addplot [color=indigo, dashed]
  table[row sep=crcr]{%
0	-18.3848705932269\\
5	-19.821750789343\\
10	-21.1377246122121\\
15	-22.2966145044631\\
20	-23.4435966252166\\
};

\end{axis}
\end{tikzpicture}%
}
    \subfloat[c][$Q$ vs. NMSE.\label{fig:epsFigQ}]{\begin{tikzpicture}
\begin{axis}[%
width=\fwidth,
scale only axis,
xmin=12,
xmax=24,
xlabel={$Q$},
ymin=-25,
ymax=-5,
xmajorgrids,
ymajorgrids,
yticklabels={},
]
\addplot [color=black, mark=triangle]
  table[row sep=crcr]{%
12	-17.474762892035\\
14	-17.7467853224144\\
16	-17.9536257238115\\
18	-17.9716641521099\\
20	-18.5217717811175\\
22	-18.4329686543145\\
24	-18.8948066749927\\
};

\addplot [color=pink, dashed, mark=star]
  table[row sep=crcr]{%
12	-9.3032657830234\\
14	-9.42499159587759\\
16	-9.50313525136367\\
18	-9.49002708817596\\
20	-9.54943885375154\\
22	-9.53319207602714\\
24	-9.51035460774085\\
};

\addplot [color=green, dashed, mark=x, mark options={solid, green}]
  table[row sep=crcr]{%
12	-5.44144439024297\\
14	-6.80918806470283\\
16	-8.12480434816373\\
18	-9.2566677721889\\
20	-10.1400981413033\\
22	-10.954591793344\\
24	-11.5612684047519\\
};

\addplot [color=red, mark=star]
  table[row sep=crcr]{%
12	-11.7505922263856\\
14	-11.7668428260734\\
16	-11.8711624958133\\
18	-11.8409707251035\\
20	-11.9040521778678\\
22	-11.8951681991242\\
24	-11.9622405354382\\
};

\addplot [color=magenta, dashed, mark=square, mark options={solid, magenta}]
  table[row sep=crcr]{%
12	-16.2008101331786\\
14	-16.2689669509485\\
16	-16.4862549936256\\
18	-16.6531642712971\\
20	-16.6070792916349\\
22	-16.8215140998527\\
24	-16.7173198062166\\
};

\addplot [color=indigo, dashed]
  table[row sep=crcr]{%
12	-19.4683726216607\\
14	-19.7118870664991\\
16	-19.7816710521393\\
18	-20.0807585163671\\
20	-20.3195146775158\\
22	-20.6469055393452\\
24	-20.6865481046938\\
};

\end{axis}
\end{tikzpicture}%
}
    \caption{NMSE performance  as a function of $P$, $\mathrm{SNR}$ and $Q$.}
    \label{fig:NMSE}
    \vspace{-.8cm}
\end{figure}
We finally compare the  complexity.
The complexity of  DS-MDT is mainly determined by the correlation operation in eq.~\eqref{eq:s5crror}, which is $\mathcal{O}(N^2g^2)$ of each path, where  $g$ denotes the number of grid points in the interval $[-\pi/N,\pi/N]$~\cite{PDMP}. 
The reference UE is required to estimate all $U^1$ paths, while the others need to estimate only $L_2^k$ paths.
Hence, the  complexity  is $\mathcal{O}(N^2 g^2 U^1)$ for the reference  UE, and only $\mathcal{O}(N^2 g^2 L_2^k)$ for the others.
For SCPD and PDMP, the complexity is $\mathcal{O}(N^2 g^2U^k)$ since all UEs are considered independent. 
Therefore, DS-MDT has a lower  complexity than the benchmarks.

\section{Conclusions \& Future works}
We developed a DS-MDT CE algorithm for RIS-aided MU-MIMO-OFDM systems. We revealed the dual-structure features of the channel parameters, and exploit them to enhance CE performance. 
We implement strategic dimensional transformations to extract distinct multi-dimensional parameters with low computational complexity.
Simulation results illustrated remarkable performance of the proposed algorithm under ideal condition and unknown number of channel paths. We remark that practical consideration like non-stationary reflectors may break the dual-structure assumption. Future works will investigate adoption of stochastic offset models--treating the offset as a random variable for joint estimation--and perturbation-aware correction mechanism for dynamic adaptation to practical environments.

\footnotesize
\bibliographystyle{IEEEtran}
\bibliography{IEEEabrv,IEEEUAF}

\normalsize
\newpage
\begin{appendices}
\section{}
Since the diagonal matrix $\mathrm{diag}({\bm h}_p^k)$ can be simplified to  
$\mathrm{diag}({\bm h}_p^k) = \sum_{l=1}^{L_2^k} \beta_{l}^k e^{-j\pi p \tau_{l,k}} \cdot \mathrm{diag}\left(\bm{a}_{N_1,N_2}(\omega_{l}^k, \psi_{l}^k)\right)$, the cascade channel ${\bm{H}}_p^k={\bm G}_p {\rm{diag}}({\bm h}_p^k)$ can be further rewritten as 
\begin{align} \label{eq:hka111}
{\bm H}_p^k = \sum_{\ell=1}^{L_1} \sum_{l=1}^{L_2^k}& \beta_{\ell} \beta_{l}^k e^{-j\pi p (\tau_{\ell} + \tau_{l}^k)} \bm{a}_{M}(\phi_{\ell}) \\ &\left[ \bm{a}_{N_1,N_2}^\top(\omega_{\ell}, \psi_{\ell}) \mathrm{diag}\left(\bm{a}_{N_1,N_2}(\omega_{l}^k, \psi_{l}^k)\right) \right] \nonumber
\end{align}

We first simplify the second line of  eq.~\eqref{eq:hka111}. 
Given that the properties of Hadamard product $\bm{a}^\top \mathrm{diag}\left(\bm b\right) = \left[ \bm{a} \odot \bm{b} \right]^\top$, we obtain
\begin{align} \label{eq:aaa111}
&\bm{a}_{N_1,N_2}^\top(\omega_{\ell}, \psi_{\ell}) \mathrm{diag}\left(\bm{a}_{N_1,N_2}(\omega_{l}^k, \psi_{l}^k)\right) \nonumber \\ 
=&\left[ \bm{a}_{N_1,N_2}(\omega_{\ell}, \psi_{\ell}) \odot \bm{a}_{N_1,N_2}(\omega_{l}^k, \psi_{l}^k) \right]^\top\\
\overset{(a)}
{=} &\left[ \bm{a}_{N_1,N_2}(\omega_{\ell} + \omega_{l}^k, \psi_{\ell} + \psi_{l}^k)\right]^\top . \nonumber
\end{align}

The step $(a)$  holds because each element of $\bm{a}_{N_1,N_2}(\cdot)$ is of the form $e^{j\pi \theta}$ as defined in eq.~\eqref{eq:array1} and eq.~\eqref{eq:array}, and element-wise multiplication adds the phases:
\begin{equation}
e^{j\pi \theta_1} \cdot e^{j\pi \theta_2} = e^{j\pi (\theta_1 + \theta_2)}
\end{equation}

Substituting eq.~\eqref{eq:aaa111} into eq.~\eqref{eq:hka111}, the cascade channel ${\bm{H}}_p^k$ can be further rewritten as 
\begin{align}
\label{eq:H1pkap}
    {\bm{H}}_p^k & = \sum\limits_{\ell = 1}^{L_1} \sum\limits_{{l=1}}^{{L^k_2}}{\beta _{{\ell}}}  \beta _{l}^k   e^{ {- j\pi p (\tau_{\ell}+\tau_{l}^k) }} {{\bm{a}}_{{{M}}}}(\phi_{ \ell } ) \cdot  \nonumber  \\ 
    & \quad \qquad \qquad \bm{a}_{N_1,N_2}\T(\omega_{\ell} + \omega_{l}^k, \psi_{\ell} + \psi_{l}^k), \nonumber  \\ 
    & = \sum\limits_{\ell = 1}^{L_1} \sum\limits_{{l=1}}^{{L^k_2}}\beta_{\ell,l}^k  e^{ {- j\pi p \tau_{\ell,l}^k  }}  {{\bm{a}}_{{{M}}}}(\phi_{ \ell } ){\bm{a}}_{N_1,N_2}\T(\omega_{\ell,l}^k,\psi_{\ell,l}^k ), \\
     & = \sum\limits_{{u=1}}^{{U^k}} \beta_{u}^k  e^{ {- j\pi p \tau_{u}^k  }}  {{\bm{a}}_{{{M}}}}(\phi_{u} ){\bm{a}}_{N_1,N_2}\T(\omega_{u}^k,\psi_{u}^k ),\nonumber
\end{align}
where$\{\phi_{\ell},\beta_{\ell,l}^k,\omega_{\ell,l}^k,\psi_{\ell,l}^k,\tau_{\ell,l}^k,\forall \ell,l,k\}$ are the cascaded parameters of the cascaded channel  and   $\{\phi_{u},\beta_{u}^k,\omega_{u}^k,\psi_{u}^k,\tau_{u}^k,\forall u,k\}$  are the the mapping parameters with  $u\triangleq(l-1)L_1+\ell, U^k = L_1L_2^k$, and  having the following mapping relationship~\cite{SCPD}
\begin{align} 
\label{eq:para111}
    \nonumber
    &\beta_{\ell,l}^k\triangleq {\beta _{{\ell}}}  \beta _{l}^k \rightarrow  \beta_{u}^k,\quad \tau_{\ell,l}^k \triangleq \tau_{\ell}+\tau_{l}^k\rightarrow\tau_{u}^k,
    \\
    &\omega_{\ell,l}^k\triangleq \omega_{\ell}+\omega_{l}^k\rightarrow \omega_{u}^k,\quad \psi_{\ell,l}^k \triangleq\psi_{\ell}+\psi_{l}^k\rightarrow\psi_{u}^k,
    \\ 
    \nonumber
    &\phi_{\ell,l}^k \triangleq \phi_{\ell}\rightarrow \phi_{u}, \quad  \forall \ell={\rm{mod}}({u},L_1),
\end{align}

\section{}
The derivations for different modes follow similar principles. 
Therefore, we  derive the Mode-1 unfolding  eq.(9) step-by-step as a representative example. 
The derivations for the Mode-2 and Mode-3 unfoldings follow similar procedures.

Recalling the defination of   tensor ${\cal \bm Y}^k \in \mathbb{C}^{P \times M \times Q}$ in eq.~\eqref{eq:Ytensor}
\begin{align}
\label{eq:Ytensor111}
    {\cal{Y}}^k &=\sum\limits_{{u=1}}^{{U^k}} { \beta_{u}^k} {\bm a}_P({\tau_{u}^k})  \circ  {\bm a}_M({ \phi_{u}}) \circ \tilde{\bm{a}}_{N_1,N_2}({\omega_{u}^k}, {\psi_{u}^k }) +{\cal \bm W }\nonumber
    \\
    &\triangleq \left[\kern-0.15em\left[ { {\bm{A}^k},{{\bm{B}}},\bm{R}^k} \right]\kern-0.15em\right]+{\cal \bm W }^k={\cal{Z}}^k+{\cal{W}}^k,
\end{align}
Its  $(p, m, q)$-th  element of the  tensor $\mathcal{Y}^k $ is:
\begin{equation}
y_{p,m,q}^k = \sum_{u=1}^{U^k} a^k_{p,u} b_{m,u} r^k_{q,u} +w_{p,m,q}^k.
\end{equation}

The  mode-1 matrix form  $\bm Y_{(1)}^k  \in \mathbb{C}^{P \times MQ}$ of ${\cal \bm Y}^k$  arranges the tensor fibers along the first dimension, with size ${P \times MQ}$. 
In detail,  its row index is 
$p$, and the column index is defined by the pair $(m,q)$ with the  mapping 
$(m,q) \rightarrow j$ 
follows $j=(m-1)Q+q$.
So the $(p,j)$-th element $[y_{(1)}^k]_{p,j}$  of  $\bm Y_{(1)}^k$  is:
\begin{align}  \label{eq:1qppx1}
[y_{(1)}^k]_{p,j} = y_{p,m,q}^k =  \sum_{u=1}^{U^k} a^k_{p,u} b_{m,u} r^k_{q,u} +w_{p,m,q}^k
\end{align}

In the other hand, the matrix $\mathbf{R}^k \odot \mathbf{B}$ has size $MQ \times U^k$ with its $u$-th column 
${\bm r^k_{u}}  \otimes { \bm b_{u}} =\begin{bmatrix} r^k_{1,u} \bm b_{u} , 
r^k_{2,u} \bm b_{u} ,... , r^k_{Q,u} \bm b_{u} , \end{bmatrix}^T,$ and its  $(j,u)$-th element ${r}^k_{q,u}{b}_{m,u}$. 
So that the $(p,j)$-th element of $\left[\mathbf{A}^k \left( \mathbf{R}^k \odot \mathbf{B} \right)^\top \right]_{p,j}$ can be calculated as

\begin{align} \label{eq:app1}
&\left[\mathbf{A}^k \left( \mathbf{R}^k \odot \mathbf{B} \right)^\top \right]_{p,j} = \sum \limits_{u=1}^{U^k} a^k_{p,u} \left[ \left( \mathbf{R}^k \odot \mathbf{B} \right)^\top \right]{u,j} \nonumber  \\
&= \sum_{u=1}^{U^k} a^k_{p,u} r^k_{q,u} b_{m,u} .
\end{align}

It can be seen that the $(p,j)$-th element of $\left[\mathbf{A}^k \left( \mathbf{R}^k \odot \mathbf{B} \right)^\top \right]_{p,j}$  in  eq.~\eqref{eq:app1} is equal to  the noiseless measurement part of the $(p,j)$-th element of the  mode-1 matrix form  $\bm Y_{(1)}^k $ in eq.~\eqref{eq:1qppx1}.
Thus:
\begin{equation}
\mathbf{Y}_{(1)}^k = \mathbf{A}^k \left( \mathbf{R}^k \odot \mathbf{B} \right)^\top + \mathbf{W}_{(1)}^k
\end{equation}

\begin{figure}[b]
    \centering    \vspace{-0.2cm}
    \includegraphics[width=0.49\textwidth]{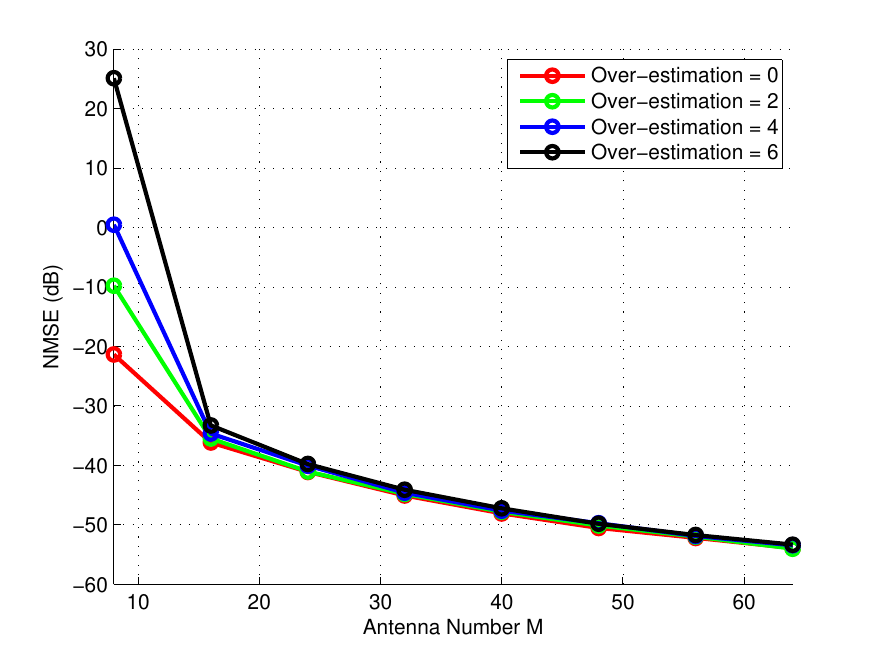}
    \caption{NMSE vs M. Two angles located at 5° and -5°. SNR=10}\label{fig:Nrang} 
    \vspace{-0.5cm}
\end{figure}

\section{}

\begin{figure}[h!]
\centering
   \subfloat[$M=8$ \label{fig:N12}]{\includegraphics[width = 0.49\textwidth]{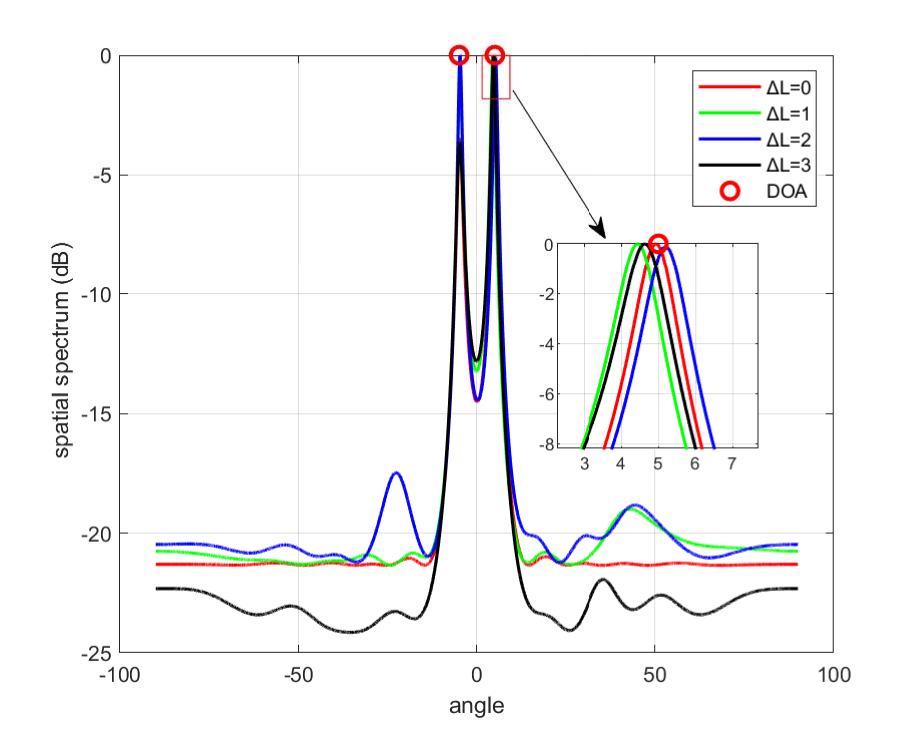}}\\
    \subfloat[$M=128$
\label{fig:N128}]{\includegraphics[width = 0.49\textwidth]{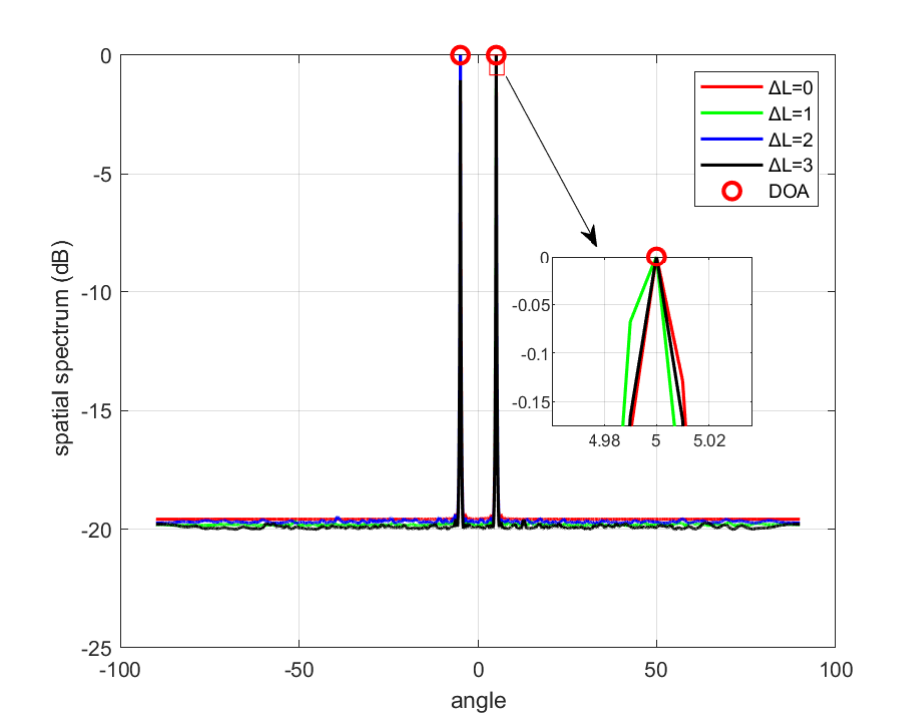}}
\caption{Impact of different 
 overestimated  source  numbers on the spatial spectrum. Two angles located at 5° and -5°. SNR=10.}
\label{fig:ASV1}
\end{figure}

As the number of antennas increases, a slight overestimation has a negligible effect on the MUSIC algorithm. The following experiments validate the above conclusions.
Fig.~\ref{fig:epsFigP} presents the NMSE as a function of the antenna count $M$.
We find that when $M \ge 16$, the NMSE of different overestimators exhibits similar performance.
Moreover, Fig.~\ref{fig:N12} and Fig.~\ref{fig:N128} illustrate the impact of overestimating the source number on the spatial spectrum for 8 and 128 antennas, respectively. When the number of antennas $M$ is 8, overestimation introduces a bias in the estimated angles. In contrast, with a larger antenna array ($M$=128), the effect of overestimation becomes insignificant.

 \end{appendices}

\end{document}